\documentclass[10pt,twocolumn,prl,aps,amssymb,amsmath,tightenlines]{revtex4}

\usepackage[dvips]{graphicx}
\usepackage{dsfont}
\newcommand{\cPT}{\ensuremath{\mathcal{PT}}}
\newcommand{\half}{\mbox{$\textstyle{\frac{1}{2}}$}}
\newcommand{\re}{{\rm e}}
\newcommand{\ri}{{\rm i}}
\newcommand{\rd}{{\rm d}}

\usepackage{xcolor}

\begin{document}
\title{Hamiltonian for the zeros of the Riemann zeta function}
\author{Carl M. Bender$^{1}$, Dorje C. Brody$^{2,3}$, Markus P. M\"uller$^{4,5}$}

\affiliation{
$^{1}$Department of Physics, Washington University, St Louis, MO 63130, USA\\
$^{2}$Department of Mathematics, Brunel University London, Uxbridge UB8 3PH,
United Kingdom\\
$^{3}$Department of Optical Physics and Modern Natural Science, St Petersburg
National Research University of Information Technologies, Mechanics and Optics, 
St Petersburg 197101, Russia \\ 
$^{4}$Departments of Applied Mathematics and Philosophy, University of Western
Ontario, Middlesex College, London, ON N6A 5B7, Canada\\
$^{5}$The Perimeter Institute for Theoretical Physics, Waterloo, ON N2L 2Y5, Canada}

\begin{abstract}
A Hamiltonian operator $\hat H$ is constructed with the property that if the eigenfunctions obey a suitable boundary condition, then the associated eigenvalues correspond to the nontrivial zeros of the Riemann zeta function. The classical limit of $\hat H$ is $2xp$, which is consistent with the Berry-Keating conjecture. While $\hat H$ is not Hermitian in the conventional sense, $\ri{\hat H}$ is ${\cal PT}$ symmetric with a broken $\cPT$ symmetry, thus allowing for the possibility that all eigenvalues of $\hat H$ are real. A heuristic analysis is presented for the construction of the metric operator to define an inner-product space, on which the Hamiltonian is Hermitian. If the analysis presented here can be made rigorous to show that ${\hat H}$ is manifestly self-adjoint, then this implies that the Riemann hypothesis holds true. 
\end{abstract}

%\pacs{03.65.Ca, 03.65.Ta, 03.65.Yz}
\maketitle

The Riemann zeta function $\zeta(z)$ is conventionally represented as the sum or
the integral
$$\zeta(z)=\sum_{k=1}^\infty\frac{1}{k^z}=\frac{1}{\Gamma(z)}\int_0^\infty
\rd t\frac{t^{z-1}}{\re^t-1}.$$
(The integral reduces to the sum if the denominator of the integrand is expanded
in a geometric series.) Both representations converge and define $\zeta(z)$ as
an analytic function when $\Re(z)>1$. These representations diverge when $z=1$
because the zeta function has a simple pole at $z=1$. Substituting $z=-2n$ 
($n=1,\,2,\,3,\,\ldots$) in the reflection formula
$$\zeta(z)=2^z\pi^{z-1}\sin(\pi z/2)\Gamma(1-z)\zeta(1-z)$$
shows that the zeta function vanishes when $z$ is a negative-even integer. These 
zeros of $\zeta(z)$ are called the {\it trivial} zeros.

The Riemann hypothesis \cite{r1} states that the {\it nontrivial} zeros of
$\zeta(z)$ lie on the line $\Re(z)=\half$. This hypothesis has
attracted much attention for over a century because there is a deep
connection with number theory and other branches of mathematics. However, the
hypothesis has not been proved or disproved. Any advance in understanding the
zeta function would be of great interest in mathematical science, whether or not
one succeeds in finally proving or falsifying the hypothesis.

In this Letter we examine the Riemann hypothesis by constructing and studying an
operator ${\hat H}$ that plays the role of a Hamiltonian. The conjectured property
of ${\hat H}$ is that its eigenvalues are exactly the imaginary parts of the nontrivial 
zeros of the zeta function. The 
idea that the imaginary parts of the zeros of $\zeta(z)$ might correspond to
the eigenvalues of a Hermitian, self-adjoint operator (assuming the validity of the 
Riemann hypothesis) is known as the {\it Hilbert-P\'olya conjecture}. Research 
into this connection has intensified following the observation that the spacings of 
the zeros of the zeta function on the line $\Re(z)=\half$ 
and the spacings of the eigenvalues of a Gaussian 
unitary ensemble of Hermitian random matrices have the same distribution 
\cite{M,r2,r3}. Berry and Keating conjectured that the classical counterpart of such a
Hamiltonian would have the form $H=xp$ \cite{r4,r5}. However, a Hamiltonian 
possessing this property has hitherto not been found (see \cite{r6} for a 
detailed account of the Berry-Keating programme and its extensions). 

We propose and consider the Hamiltonian
\begin{eqnarray}
{\hat H}=\frac{{\mathds 1}}{{\mathds 1}-\re^{-{\rm i}{\hat p}}}\left({\hat x}{\hat p}+
{\hat p}{\hat x}\right)({\mathds 1}-\re^{-{\rm i}{\hat p}}). 
\label{e1}
\end{eqnarray}
Our main findings are as follows: (i) The non-Hermitian Hamiltonian ${\hat H}$ 
in (\ref{e1}) formally satisfies the conditions of the Hilbert-P\'olya conjecture. 
That is, if the eigenfunctions of ${\hat H}$ are required 
to satisfy the boundary condition $\psi_n(0)=0$ for all $n$, then the eigenvalues 
$\{E_n\}$ have the property that $\{\half(1-\ri E_n)\}$ are the nontrivial zeros of the 
Riemann zeta function. 
(ii) The Hamiltonian ${\hat H}$ reduces to the classical Hamiltonian 
$H=2xp$ when ${\hat x}$ and ${\hat p}$ commute, in agreement with the 
Berry-Keating conjecture. We derive the corresponding boundary 
condition that leads to the quantization of the Berry-Keating Hamiltonian 
${\hat h}^{\rm BK}={\hat x}\,{\hat p}+{\hat p}\,{\hat x}$. 
(iii) Although $\hat H$ is not Hermitian, $\ri{\hat H}$ is $\cPT$ symmetric; 
that is, $\ri{\hat H}$ is invariant under parity-time reflection (in the sense 
to be defined), which means that the eigenvalues of $\ri{\hat H}$ are 
either real or else occur in complex-conjugate pairs. If $\ri{\hat H}$ has {\it 
maximally broken} $\cPT$ symmetry; that is, if all of its eigenvalues are 
{\it pure-imaginary complex-conjugate pairs}, then the eigenvalues of 
${\hat H}$ are real and the Riemann hypothesis follows. 
(iv) While ${\hat H}$ is not Hermitian (symmetric) 
with respect to the conventional ${\cal L}^2$ inner product, we introduce 
an alternative inner product such that 
$\langle {\hat H}\varphi,\psi\rangle = \langle \varphi,{\hat H}\psi\rangle$ 
for all $\varphi(x)$ and $\psi(x)$ belonging to the linear span of the 
eigenstates of ${\hat H}$. 
(v) If the Riemann hypothesis is correct, 
then the eigenvalues of ${\hat H}$ are nondegenerate, and conversely 
if there are nontrivial roots of $\zeta(z)$ for which $\Re(z)\neq\frac{1}{2}$ 
then the corresponding eigenvalues and eigenstates are both degenerate. 

\noindent{\it Preliminaries}. The Hamiltonian ${\hat H}$ in (\ref{e1}) is a 
similarity transformation of the formally Hermitian local 
Hamiltonian ${\hat x}{\hat p}+{\hat p}{\hat x}$ via the nonlocal operator 
${\hat\Delta}:={\mathds 1}-\re^{-{\rm i}{\hat p}}$. We must therefore
identify properties of the operators ${\hat\Delta}$ and ${\hat \Delta}^{-1}$. 
We work in units for which $\hbar=1$, so the momentum operator is 
${\hat p}=-\ri\partial_x$. Thus, $\re^{-{\rm i}{\hat p}}$ is a shift operator if it 
acts on functions $f(x)$ that have a Taylor series about $x$ with radius of
convergence greater than one. In this case ${\hat \Delta}$ is a difference
operator:
\begin{eqnarray}
{\hat\Delta}f(x)=f(x)-f(x-1). 
\label{e2}
\end{eqnarray}
Because ${\hat\Delta}$ annihilates unit-periodic functions, it does not 
have an inverse in the space of all smooth functions. However, we shall 
be interested in functions that vanish as $x\to\infty$. With this in mind, 
by taking a series expansion of $({\mathds 1}-\re^{-{\rm i}{\hat p}})^{-1}$ 
we may define ${\hat\Delta}^{-1}$ as (cf. \cite{rZ}) 
\begin{eqnarray}
{\hat\Delta}^{-1} f(x) = \frac{{\mathds 1}}{\ri{\hat p}} \,
\sum_{n=0}^\infty B_n \frac{(-\ri {\hat p})^n}{n!} f(x),
\label{e3} 
\end{eqnarray}
where $\{B_k\}$ are the Bernoulli numbers \cite{r9}, with the 
convention that $B_1=-\frac{1}{2}$. 
For some functions $f(x)$ this formal series diverges but it is Borel summable. 
The 
operator $(\ri{\hat p})^{-1}$ is interpreted as an integral operator with boundary 
at infinity: 
\[ 
\frac{{\mathds 1}}{\ri{\hat p}} \, g(x) = \int_\infty^x \rd t \, g(t) . 
\] 
Then ${\hat\Delta}^{-1}$ defined in (\ref{e3}) has the property that if $f(x)$ 
vanishes at infinity, then we have ${\hat\Delta}^{-1}{\hat\Delta}f(x)=f(x)$. 

\noindent{\it Eigenfunctions and eigenvalues}. The solutions to the eigenvalue 
differential equation ${\hat H}\psi=E\psi$ are given in terms of the 
Hurwitz zeta function $\psi_z(x)=-\zeta(z,x+1)$ on the positive half line 
${\mathds R}^+$ (the negative sign is our convention), with eigenvalues 
$\ri(2z-1)$. To see this, we multiply the eigenvalue equation ${\hat H}\psi=
E\psi$ on the left by $\hat\Delta$. This gives a first-order linear differential 
equation $({\hat x}{\hat p}+{\hat p}{\hat x}){\hat\Delta}\psi=E{\hat\Delta}\psi$ 
for the function ${\hat\Delta}\psi$, whose solution is unique and is given by 
${\hat\Delta}\psi=x^{-z}$ for some $z\in{\mathds C}$, up to a 
multiplicative constant. To proceed, let us calculate 
\begin{eqnarray} 
{\hat\Delta}^{-1} x^{-z} &=& \frac{{\mathds 1}}{\ri{\hat p}} \, 
\sum_{n=0}^\infty B_n \frac{(-\ri {\hat p})^n}{n!} (\ri{\hat p}) \frac{x^{1-z}}{1-z} 
\nonumber \\ &=& \frac{1}{1-z} \sum_{n=0}^\infty B_n \frac{(-\ri {\hat p})^n}{n!} 
x^{1-z} . \nonumber 
\end{eqnarray}
Since $\ri {\hat p}=\partial_x$ and
$\partial_x^n \, x^\mu = [\Gamma(\mu+1)/\Gamma(\mu-n+1)]\, x^{\mu-n}$, 
we set $\mu=1-z$ to obtain the asymptotic series 
\begin{eqnarray}
{\hat\Delta}^{-1} x^{-z} \sim 
\frac{\Gamma(2-z)}{1-z} \sum_{n=0}^\infty B_n \frac{(-1)^n}{n!} 
\frac{x^{1-z-n}}{\Gamma(2-z-n)} ,
\label{eqn}
\end{eqnarray}
which is valid in the limit as $x\to\infty$. To obtain the Borel sum \cite{cmb} of 
the series, we use the integral representation 
\[
\frac{1}{\Gamma(2-z-n)}=\frac{1}{2\pi\ri}\int_C\rd u\,\re^u\,u^{n+z-2},
\]
where $C$ denotes a Hankel contour that encircles the negative-$u$
axis in the positive orientation \cite{r9}. Hence,
\begin{eqnarray} 
{\hat\Delta}^{-1} x^{-z} &=& \frac{\Gamma(1-z)}{2\pi\ri} \, x^{1-z}\! \int_C\rd u\, 
\re^u\,u^{z-2} \sum_{n=0}^\infty B_n \frac{(-u/x)^n}{n!} \nonumber \\ &=& 
\frac{\Gamma(1-z)}{2\pi\ri} \, x^{-z} \! \int_C\rd u\, 
\frac{\re^u u^{z-1}}{1-\re^{-u/x}} . \nonumber 
\end{eqnarray}
Finally, we let $u/x=t$ and get
\[
{\hat\Delta}^{-1} x^{-z} = 
\frac{\Gamma(1-z)}{2\pi\ri} \int_C\rd t\, 
\frac{\re^{xt} t^{z-1}}{1-\re^{-t}} , 
\] 
which we recognise as the negative of the integral representation for the 
Hurwitz zeta function \cite{r9}. 
(An analogous result was obtained in a different context in \cite{r8}.) 
It follows that $\psi_z(x)=-\zeta(z,x+1)$ up to an additive 
unit-periodic function, but ${\hat H}\psi=E\psi$ implies that the periodic 
function must be identically zero. We thus deduce that 
$\psi_z(x)=-\zeta(z,x+1)$ is the solution to the eigenvalue differential 
equation with eigenvalue $\ri(2z-1)$: 
$${\hat H}\psi_z(x)={\hat\Delta}^{-1}\left({\hat x}{\hat p}+{\hat p}{\hat x}
\right)x^{-z}=\ri(2z-1)\psi_z(x).$$

Next, we impose the boundary condition that $\psi_z(0)=0$ on the 
class of functions $\psi_z(x)$ that satisfy the eigenvalue differential 
equation. This yields a countable set of eigenfunctions of 
${\hat H}$. (Since ${\hat H}$ is similar to a first-order differential operator, 
we impose just one boundary condition.) The choice of the 
boundary condition $\psi_z(0)=0$, as discussed below, is motivated 
by our requirement that ${\hat p}$ should be symmetric. 
Because $-\psi_z(0)=\zeta(z)$ is the
Riemann zeta function, the boundary condition that we have used 
implies that $z$ must belong to
the {\it discrete} set of zeros of $\zeta(z)$. 

The zeros of the Riemann zeta function 
may be either trivial or nontrivial. It follows from (\ref{eqn}) that for the trivial 
zeros $z=-2n$ ($n=1,\,2,\,3,\,\ldots$) we have 
$\psi_z(x)=-B_{2n+1}(x+1)/(2n+1)$, where $B_n(x)$ is a
Bernoulli polynomial \cite{r9}. In this case $|\psi_z(x)|$ grows like 
$x^{2n+1}$ as $x\to\infty$. For the nontrivial zeros $\psi_z(x)$ oscillates and 
$|\psi_z(x)|$ grows sublinearly. In particular, it follows from (\ref{eqn}) that 
for large $x$ we have $\psi_z(x)\approx x^{1-z}/(1-z)$. Thus, for the 
trivial zeros ${\hat\Delta}\psi_z(x)$ blows up, but for the nontrivial zeros 
${\hat\Delta}\psi_z(x)$ goes to zero as $x\to\infty$. The eigenstates associated 
with the trivial zeros violate the orthogonality relation discussed below and the 
eigenstates associated with the nontrivial zeros do not. These indicate that the 
eigenstates associated with the trivial zeros do not 
belong to the domain of ${\hat H}$. Therefore, under the boundary condition 
$\psi(0)=0$, the $n$th eigenstate of the Hamiltonian (\ref{e1}) is 
$\psi_n(x)=-\zeta(z_n,x+1)$; the eigenvalues
$E_n=\ri(2z_n-1)$
are discrete and $z_n=\frac{1}{2}(1-\ri E_n)$ are the nontrivial zeros of the
Riemann zeta function. The Riemann hypothesis is valid if and only if 
these eigenvalues are real.

The analysis above establishes a complex extended version of the Berry-Keating
conjecture \cite{rX}. We are not able to prove that the eigenvalues of ${\hat H}$ 
are real; nevertheless, in what follows we present a heuristic analysis that
suggests that the eigenvalues are real. Specifically, we first investigate 
symmetry properties of ${\hat H}$, which shows that $\ri {\hat H}$ is 
$\cPT$-symmetric and ${\hat H}$ is pseudo-Hermitian. 
This allows us to obtain a quantization of the Berry-Keating Hamiltonian 
${\hat h}^{\rm BK}={\hat x}\,{\hat p}+{\hat p}\,{\hat x}$ that is isospectral to ${\hat H}$. 
We then make use 
of the biorthogonality properties of the eigenstates of ${\hat H}$ to introduce an 
inner product which makes ${\hat H}$ Hermitian.

\noindent{\it Relation to pseudo-Hermiticity}. To gain some intuition about the
reality of the eigenvalues of the Hamiltonian, we remark first that $\ri{\hat H
}$ is $\cPT$ symmetric \cite{r10,r11} in the following sense. Under conventional
parity-time reflection, if $\hat p$ is a momentum and $\hat x$ is a coordinate,
we have $\cPT:\,({\hat x},{\hat p})\longrightarrow(-{\hat x},{\hat p})$.
However, we consider instead the variables where the roles of position ${\hat x}$ 
and momentum ${\hat p}$
are interchanged \cite{r12}. We then define parity-time reflection as $\cPT:\,
({\hat x},{\hat p})\longrightarrow({\hat x},-{\hat p})$. Therefore, since $\cPT:
\,\ri\longrightarrow-\ri$, we deduce that $\ri{\hat H}$ is invariant under this
modified $\cPT$ reflection. It follows that the eigenvalues of $\ri{\hat H}$ are
either real (if the $\cPT$ symmetry is \textit{unbroken} in the sense that the 
associated eigenstates are also eigenstates of $\cPT$), or else they form 
complex-conjugate pairs (if the $\cPT$ symmetry is \textit{broken} in the sense that 
the associated eigenstates are not eigenstates of $\cPT$). If the $\cPT$ symmetry 
is maximally
broken for $\ri{\hat H}$, then the eigenvalues of ${\hat H}$ would be real, and
the Riemann hypothesis would hold. In our case, since $\cPT\psi_n(x)=\psi_{-n}(x
)$, the $\cPT$ symmetry is indeed broken for all complex values of $z_n$. 
(For the trivial zeros the $\cPT$ symmetry is unbroken.) 

Let us now {\it assume} that the momentum operator ${\hat p}$ is Hermitian 
(symmetric); that is, the action of ${\hat p}^\dag$ agrees with that of ${\hat p}$ 
on the domain of ${\hat H}$. Here $\dag$ 
denotes the adjoint with respect to the standard inner product on 
${\cal L}^2({\mathds R}^+)$.
Then the Hermitian 
adjoint of ${\hat H}$ is
\begin{eqnarray}
{\hat H}^\dag=({\mathds 1}-\re^{{\rm i}{\hat p}})\left({\hat x}{\hat p}+
{\hat p}{\hat x}\right)\frac{{\mathds 1}}{{\mathds 1}-\re^{{\rm i}{\hat p}}}.
\label{e5}
\end{eqnarray}
Therefore, if we define the operator ${\hat\eta}$ according to
$${\hat\eta}=\sin^2\half{\hat p},$$
which is nonnegative, bounded, and Hermitian under the assumption, we get 
${\hat H}^\dagger={\hat\eta}{\hat H}{\hat\eta}^{-1}$, i.e. 
${\hat H}$ is \textit{pseudo-Hermitian} in the sense of \cite{r13}.
Assuming that ${\hat p}$ is Hermitian, there exists an associated Hermitian
Hamiltonian ${\hat h}$ obtained by conjugating ${\hat H}$ with 
an operator ${\hat\rho}$ satisfying ${\hat\rho}^\dagger{\hat\rho}={\hat\eta}$, 
that is, ${\hat\rho}{\hat H}{\hat\rho}^{-1}={\hat h}$. Letting 
${\hat\rho}=\sin\half{\hat p}$, we obtain 
${\hat h}={\hat x}\,{\hat p}+{\hat p}\,{\hat x}+\hbar{\hat p}$. 
We include Planck's constant $\hbar$ explicitly here because it
indicates that the linear momentum term is a {\it quantum anomaly}; this term
vanishes in the classical limit $\hbar\to0$ \cite{r12}. 
Alternatively, by letting ${\hat\rho}={\hat\Delta}$ we obtain the Berry-Keating 
Hamiltonian ${\hat h}^{\rm BK}={\hat x}\,{\hat p}+{\hat p}\,{\hat x}$, whose 
eigenstates are $\phi_z^{\rm BK}(x)=x^{-z}$. 

The associated Hamiltonian ${\hat h}$ is unique up to unitary transformations, 
so there are infinitely many formally Hermitian Hamiltonians that are 
similar to ${\hat H}$ \cite{rX}. If both ${\hat\eta}$ and 
${\hat\eta}^{-1}$ are positive, bounded, and Hermitian, then the 
Hamiltonians ${\hat H}$ and ${\hat h}$ are isospectral \cite{r14}. Assuming 
that ${\hat p}$ is Hermitian, these operators are indeed Hermitian and 
nonnegative, but ${\hat\eta}^{-1}$ is not bounded. Nevertheless, we 
can show by a direct calculation that ${\hat H}$ and ${\hat h}$ are in fact 
isospectral. 
Furthermore, since the map from the eigenstates $\{\psi_n(x)\}$ of ${\hat H}$ to 
the eigenstates $\{\phi_n(x)\}$ of ${\hat h}$ is governed by ${\hat\rho}$, we 
can identify the quantisation condition for the eigenstates of the 
associated Hamiltonians explicitly by using the relation 
$2\ri\sin\half{\hat p}\,\psi_z(x)=\psi_z(x+\frac{1}{2})-\psi_z(x-\frac{1}{2})$. 
For the Berry-Keating Hamiltonian, the condition $\psi_z(0)=0$ leads to 
\[ 
\lim_{x\to0}\left[\phi_z^{\rm BK}(x)-\zeta(z,x-1)\right]=0, 
\] 
or equivalently, $\lim_{x\to1}\phi_z^{\rm BK}(x)=-\lim_{x\to1}\zeta(z,x+1)$. 

\noindent{\it Biorthogonal states}. Let us proceed under the assumption that
${\hat p}$ is Hermitian. Because ${\hat H}$ is not Hermitian, its eigenstates
$\{\psi_n(x)\}$ are not orthogonal. Nevertheless, by considering the eigenstates
$\{{\tilde\psi}_n(x)\}$ of ${\hat H}^\dagger$ we obtain a biorthogonal set of
eigenstates \cite{r14}, provided that ${\hat H}^\dagger$ is the Hermitian 
adjoint of ${\hat H}$. Bearing in mind that ${\hat\Delta}^\dag$ is the forward
difference operator, a calculation shows that ${\tilde\psi}_n(x)=x^{-z_n}-(x+1)^{-z_n}$
and that ${\hat H}^\dagger{\tilde\psi}_n(x)=\ri(2z_n-1){\tilde\psi}_n(x)$. Using $\{{\tilde\psi}_n(x)
\}$, we introduce an inner product on the space of functions spanned by
$\{\psi_n(x)\}$ as follows. For any $\psi(x)=\sum_n c_n\psi_n(x)$ we define 
its associated state by ${\tilde\psi}(x)=\sum_n c_n
{\tilde\psi}_n(x)$. The inner product of a pair of such functions $\psi(x)$ and $\varphi
(x)$ is then defined by $\langle\varphi,\psi\rangle=\langle{\tilde\varphi}|
\psi\rangle:=\int_0^\infty\overline{{\tilde\varphi}(x)}\psi(x)\rd x$. 
Alternatively stated, since
${\tilde\varphi}(x)={\hat\eta}\varphi(x)$, we have $\langle\varphi,\psi\rangle=
\langle\varphi|{\hat\eta}|\psi\rangle$; that is, the positive Hermitian operator
${\hat\eta}$ plays the role of the metric (or equivalently the $\mathcal{CP}$
operator \cite{r15}).

For ${\hat H}$ in (\ref{e1}) the inner-product space constructed above is not 
a Hilbert space because, as we will see, the elements of the vector space have 
infinite norm. However, the elements of $\{\psi_n(x)\}$ and those of 
$\{{\tilde\psi}_n(x)\}$ are biorthogonal provided that $\{z_n\}$ belongs to the 
nontrivial zeros of the Riemann zeta function. To see this, let us consider the 
inner product $\langle{\tilde\psi}_m|\psi_n\rangle$. Observing that 
\[ 
{\tilde\psi}_m(x)={\hat\Delta}^\dagger{\hat\Delta}\psi_n(x) = {\hat\Delta}^\dagger 
{\hat\Delta} {\hat\Delta}^{-1} x^{-z_m} = {\hat\Delta}^\dagger x^{-z_m}, 
\] 
and recalling that $\psi_n(x)={\hat\Delta}^{-1}x^{-z}$, we find that 
\begin{eqnarray}
\langle{\tilde\psi}_m|\psi_n\rangle&=&\int_0^\infty\rd x\,
x^{-{\bar z}_m} {\hat\Delta} {\hat\Delta}^{-1} x^{-z_n} 
\nonumber\\
&=&\int_0^\infty\rd x\,x^{-1+{\rm i}(E_n-{\bar E}_m)/2}.
\label{e7}
\end{eqnarray}
Thus, if ${\bar E}_m=E_m$ (that is, if the Riemann hypothesis is correct), 
then (\ref{e7}) is a Dirac delta function $4\pi\delta(E_n-E_m)$. It follows 
that for $m\neq n$ we have 
\begin{eqnarray}
\langle{\tilde\psi}_m|\psi_n\rangle = 0 
\label{e8}
\end{eqnarray}
in the distributional sense, as required by the biorthogonality condition. 
In contrast, for the trivial zeros, the integral (\ref{e7}) diverges too rapidly 
to be interpreted as a tempered distribution.  

In terms of the inner product introduced above, and assuming that ${\hat p}$ 
is Hermitian (symmetric), we find, using ${\hat\Delta}^\dagger{\hat\Delta}
={\hat\eta}$, that  
\begin{eqnarray}  
\langle {\hat H}\varphi,\psi\rangle &=& \int_0^\infty \rd x \, {\bar\varphi}(x) 
{\hat\Delta}^\dagger ({\hat x}\,{\hat p}+{\hat p}\,{\hat x}) 
({\hat\Delta}^\dagger)^{-1} {\hat\Delta}^\dagger {\hat\Delta} \psi(x) 
%\nonumber \\ &=& \int_0^\infty \rd x \, {\bar\varphi}(x) 
%{\hat\Delta}^\dagger ({\hat x}\,{\hat p}+{\hat p}\,{\hat x}) {\hat\Delta} \psi(x) 
\nonumber \\ &=& \int_0^\infty \rd x \, {\bar\varphi}(x) 
{\hat\Delta}^\dagger {\hat\Delta} {\hat\Delta}^{-1} ({\hat x}\,{\hat p}+
{\hat p}\,{\hat x}) {\hat\Delta} \psi(x) \nonumber \\ &=&
\langle \varphi,{\hat H}\psi\rangle .
\nonumber 
\end{eqnarray} 
This shows that, from the assumption that ${\hat p}$ is Hermitian, we may 
conclude that ${\hat H}$ is Hermitian (symmetric) with respect 
to the new inner product. 

As a further consequence of (\ref{e7}) and (\ref{e8}), if the Riemann 
hypothesis is true, then the eigenvalues of ${\hat H}$ are nondegenerate. 
Conversely, if the Riemann hypothesis is false, then the eigenstates of ${\hat H}$ 
that correspond to nontrivial zeros for which $\Re(z)\neq\frac{1}{2}$ coalesce to 
give rise to Jordan block structures in the Hamiltonian. This follows from the fact 
that at such complex degeneracies (often referred to as \textit{exceptional points}), 
the eigenstates satisfy the so-called self-orthogonality condition 
$\langle{\tilde\psi}_n|\psi_n\rangle=0$. These findings may have an implication on 
whether the zeros of $\zeta(z)$ are simple: It is known that if the Riemann hypothesis 
holds true, then at least $19/27$ of the nontrivial zeros are simple \cite{r16}. 
However, if there exists a one-to-one correspondence between the boundary 
condition on the eigenstates of ${\hat H}$ and the secular equation for the eigenvalues 
of ${\hat H}$, then it follows that the validity of the Riemann hypothesis implies that 
all roots are simple, and conversely any nontrivial zero of $\zeta(z)$ for which $\Re(z)\neq
\frac{1}{2}$ cannot be simple.

\noindent{\it Boundary condition revisited}. For finite-dimensional
nondegenerate matrices, the biorthogonality relation (\ref{e8}) implies that
${\hat H}^\dagger$ defined in (\ref{e5}) is the Hermitian adjoint of
${\hat H}$. However, in infinite-dimensional vector spaces the completeness of
the states $\{\psi_n(x)\}$ is required to arrive at this conclusion. 
Nevertheless, the relation (\ref{e8}) suggests that our Hermiticity assumption 
of ${\hat p}$ is valid, making ${\hat h}$ manifestly Hermitian. 

Encouraged by this observation, we ask whether the momentum operator ${\hat p}$
is Hermitian (symmetric) on the inner-product space defined above. Because 
$[{\hat p},{\hat\eta}]=0$, the Hermiticity of ${\hat p}$ on $\langle\cdot,\cdot\rangle$
follows if the boundary terms vanish under an integration by parts when the elements 
of $\{\psi_n(x)\}$ and those of $\{{\tilde\psi}_n(x)\}$ are paired. Note that ${\tilde\psi}_n(x)$
diverges at $x=0$, so $\psi_n(x)$ must vanish sufficiently fast at $x=0$ to 
ensure the vanishing of the boundary terms. 
[The divergence of $\{\psi_n(x)\}$ at $x=\infty$ is compensated by the vanishing of 
$\{{\tilde\psi}_n(x)\}$ as $x\to\infty$.] 
One can verify that imposing $\psi_n(0)=0$ is sufficient to guarantee 
the vanishing of the boundary term at the origin. Thus, the
Hermiticity of ${\hat p}$ on $\langle\cdot,\cdot\rangle$ follows from 
the boundary condition $\psi_n(0)=0$.

\noindent{\it Relation to quantum mechanics}. Since the operator ${\hat H}$ is 
a function of the canonical variables $({\hat x},{\hat p})$, we have referred to
it as a Hamiltonian. However, the connection of this Hamiltonian to physical
systems is at best tenuous because the eigenstates of ${\hat H}$ in our
inner-product space are not normalizable. This is not a concern
for our analysis but in quantum mechanics normalizability is required for a
probabilistic interpretation.

A possible way of making a connection to quantum theory is to introduce a
regularization scheme, for example, by letting $x\in[\Lambda^{-1},\Lambda]$,
renormalizing the states according to $\psi_n(x)\to(\ln\Lambda)^{-1/2}\psi_n(x
)$, and then taking the limit $\Lambda\to\infty$. Interestingly, the expectation
value of the position operator ${\hat\rho}^{-1}{\hat x}{\hat\rho}$ in
the state $\psi_n(x)$ for any $n$ in the renormalized theory is $\Lambda/\ln
\Lambda$, which for large $\Lambda$ gives the leading term in the counting of
prime numbers smaller than $\Lambda$. 

\noindent{\it Discussion}. We have presented a formal argument showing that 
the eigenvalues of the Hamiltonian ${\hat H}$ in (\ref{e1}), whose classical 
limit is $2xp$, correspond to the nontrivial zeros of the Riemann zeta function. 
Identifying the domain of ${\hat H}$ remains a difficult and open problem. 
We hope that further analysis of the properties of ${\hat H}$, such as 
identifying its domain and establishing its self-adjointness, 
will prove the reality of the eigenvalues, and thus the veracity of the 
Riemann hypothesis. 
The possibility of extending 
the Hilbert-P\'olya program to non-Hermitian $\cPT$-symmetric operators 
has been noted \cite{rY}. We hope that our findings will 
significantly boost research in this direction. The fact that $\ri{\hat H}$ is 
$\cPT$ symmetric, with a broken $\cPT$ symmetry, offers a fresh and optimistic 
outlook.

DCB thanks D. Blasius and C. Hughes for comments and 
the Russian Science Foundation for support (project
16-11-10218). MPM thanks 
D. Schleicher for discussions. 
MPM is supported in part by the Canada Research Chairs
program. Research at Perimeter Institute 
is supported by the Government of Canada through Innovation, Science and Economic 
Development Canada and by the Province of Ontario through the Ministry of Research, 
Innovation and Science.


\begin{thebibliography}{999}

\bibitem{r1} B.~Riemann, Ueber die Anzahl der Primzahlen unter einer gegebenen
Gr\"osse. {\em Monatsberichte der Berliner Akademie} (1859).

\bibitem{M} H.~L.~Montgomery,  
The pair correlation of zeros of the zeta function. 
{\em Analytic number theory}. Proc. Sympos. Pure Math. 
\textbf{XXIV}, Providence, R.I.: American Mathematical Society, 
pp. 181-193 (1973). 

\bibitem{r2} A.~M.~Odlyzko, 
On the distribution of spacings between zeros of the zeta function. 
{\em Mathematics of Computation} {\bf 48}, 273-308 (1987).

\bibitem{r3} M.~V. Berry, Riemann's zeta function: A model for quantum chaos?
In {\em Quantum Chaos and Statistical Nuclear Physics}. T.~H.~Seligman and
H.~Nishioka (eds), Lecture Notes in Physics, {\bf 263} (Springer-Verlag, New
York, 1986).

\bibitem{r4} M. V. Berry and J. P. Keating, H=xp and the Riemann zeros. 
In {\em Supersymmetry and Trace Formulae: Chaos and Disorder}, edited by 
I.V.~Lerner {\it et al}. (Kluwer Academic/Plenum: New York, 1999).

\bibitem{r5} A.~Connes, Trace formula in noncommutative geometry and the zeros
of the Riemann zeta function. 
{\em Selecta Mathematica, New Series} {\bf 5}, 29-106 (1999). 

\bibitem{r6} G.~Sierra, The Riemann zeros as spectrum and the Riemann
hypothesis. arXiv:1601.01797 (2016). 

\bibitem{rZ} \"E.~Delabaere, Ramanujan's summation. 
{\em Algorithms Sem.} \textbf{2001-2002}, 83 (2003). 

\bibitem{r9} F. W. J. Olver, D. M. Lozier, R. F. Boisvert, and C. W. Clark,
{\it NIST Handbook of Mathematical Functions} (Cambridge University Press,
Cambridge, 2010). 

%\bibitem{r7} Zeta-function summation is used to solve physical problems
%involving summation over modes. One example is the determination of the
%Casimir force; see K. Milton, {\it The Casimir Effect: Physical
%Manifestations of Zero-Point Energy} (World Scientific, Singapore, 2001).

\bibitem{cmb} C.~M.~Bender and S.~A.~Orszag, 
{\em Advanced Mathematical Methods for Scientists and Engineers} 
(New York: McGraw-Hill, 1978). 

\bibitem{r8} M.~M\"uller and D.~Schleicher, 
How to add a non-integer number of terms, and how to produce unusual infinite
summations. {\em J. Comp. App. Math.}~{\bf 178}, 347-360 (2005); 
Fractional sums and Euler-like identities. {\em Ramanujan J.} \textbf{21}, 
123-143 (2010); 
How to add a noninteger number of terms: from axioms to new identities. 
{\em Amer. Math. Month.}~{\bf 118}, 136-152 (2011).

\bibitem{rX} One can extend ${\hat H}$ to a 
one-parameter family of Hamiltonians ${\hat H}_\varepsilon$ by the 
replacement ${\hat\Delta}\to{\hat\Delta}_\varepsilon=\varepsilon^{-1} 
( {\mathds 1}-\re^{-{\rm i}\varepsilon{\hat p}})$. A 
calculation shows that the eigenstates $\psi_z^\varepsilon(x)$ of 
${\hat H}_\varepsilon$ take 
the form $\psi_z^\varepsilon(x) \propto - \zeta(z,1+x/\varepsilon)$ with 
eigenvalue $\ri (2z-1)$. In the limit $\varepsilon\to0$ 
we obtain the Hamiltonian ${\hat p}^{-1}({\hat x}\,{\hat p}+{\hat p}\,
{\hat x}){\hat p}$ with eigenstate $x^{1-z}$. 

\bibitem{r10} C.~M.~Bender, Making sense of non-Hermitian Hamiltonians. 
{\em Rep.~Prog.~Phys.}~{\bf 70}, 947-1018 (2007). 
 
\bibitem{r11} D.~C.~Brody, Consistency of PT-symmetric quantum mechanics. 
{\em J.~Phys.~A: Math.~Theor.}~{\bf 49}, 10LT03 (2016).

\bibitem{r12} C.~M.~Bender, D.~C.~Brody, J.-H.~Chen, H.~F.~Jones, 
K.~A.~Milton, and M.~C.~Ogilvie, 
Equivalence of a complex PT-Symmetric quartic Hamiltonian and a Hermitian
quartic Hamiltonian with an anomaly. 
{\em Phys.~Rev.~D} {\bf 74}, 025016 (2006).

\bibitem{r13} G.~W.~Mackey, 
{\em Commutative Banach Algebras}. 
(Instituto de Matematica pura e Aplicada do Conselho Nacional de Pesquisa, 
Rio De Janeiro, 1959).

\bibitem{r14} D.~C.~Brody, 
Biorthogonal quantum mechanics. 
{\em J. Phys. A: Math. Theor.}~{\bf 47}, 035305 (2014).

\bibitem{r15} C.~M.~Bender, D.~C.~Brody, and H.~F.~Jones,
Complex extension of quantum mechanics. 
{\em Phys.~Rev.~Lett.}~{\bf 89}, 270401 (2002).

\bibitem{r16} Bui,~H.~M. and Heath-Brown,~D.~R. 
On simple zeros of the Riemann zeta-function. 
{\em Bull. Lonson Math. Soci} \textbf{45}, 953-961 (2013). 

\bibitem{rY} Z.~Ahmed and S.~R.~Jain, 
A pseudo-unitary ensemble of random matrices, PT-symmetry and the
Riemann hypothesis. {\em Mod. Phys. Lett.} A\textbf{21}, 331-338 
(2006) 

\end{thebibliography}
\end{document}